\documentclass[12pt,tightenlines,nofootinbib]{revtex4-2}
\usepackage{amsfonts,amsmath,amssymb}
\usepackage{txfonts}
\usepackage{graphicx,color}
\usepackage{wrapfig,lipsum}
\usepackage{floatrow,float,booktabs}
\usepackage{epstopdf}
\usepackage{hyperref}
\usepackage[normalem]{ulem}
\usepackage{enumitem}
\usepackage{braket}
\usepackage{mathrsfs,mathtools}

\usepackage{cancel}
\usepackage{lineno}

\begin{document}

\title{Theta oscillons in behaving rats}
\author{M. S. Zobaer$^{1}$, N. Lotfi$^{1}$, C. M. Domenico$^{2}$, C. Hoffman$^{1}$, L. Perotti$^{3}$,
	D. Ji$^{2}$, Y. Dabaghian$^{1*}$}
\affiliation{
$^1$Department of Neurology, The University of Texas Health Science Center at Houston, Houston, TX 77030\\
$^2$Department of Neuroscience, Baylor College of Medicine, Houston, TX 77030,\\
$^3$Department of Physics, Texas Southern University, 3100 Cleburne Ave., Houston, Texas 77004,\\
$^{*}$e-mail: yuri.a.dabaghian@uth.tmc.edu}
\vspace{17 mm}
\date{\today}
\vspace{17 mm}
\begin{abstract}
	Recently discovered constituents of the brain waves---the \textit{oscillons}---provide high-resolution
	representation of the extracellular field dynamics. Here we study the most robust, highest-amplitude
	oscillons that manifest in actively behaving rats and generally correspond to the traditional 
	$\theta$-waves. We show that the resemblances between $\theta$-oscillons and the conventional $\theta$-waves
	apply to the ballpark characteristics---mean frequencies, amplitudes, and bandwidths. In addition, both
	hippocampal and cortical oscillons exhibit a number of intricate, behavior-attuned, transient properties
	that suggest a new vantage point for understanding the $\theta$-rhythms' structure, origins and functions.
	We demonstrate that oscillons are frequency-modulated waves, with speed-controlled parameters, embedded
	into a noise background. We also use a basic model of neuronal synchronization to contextualize and to 
	interpret the observed phenomena. In particular, we argue that the synchronicity level in physiological
	networks is fairly weak and modulated by the animal's locomotion.
\end{abstract}
\maketitle

\newpage

\section{INTRODUCTION}
\label{sec:intro}

\begin{figure}[h]
	\includegraphics[scale=0.75]{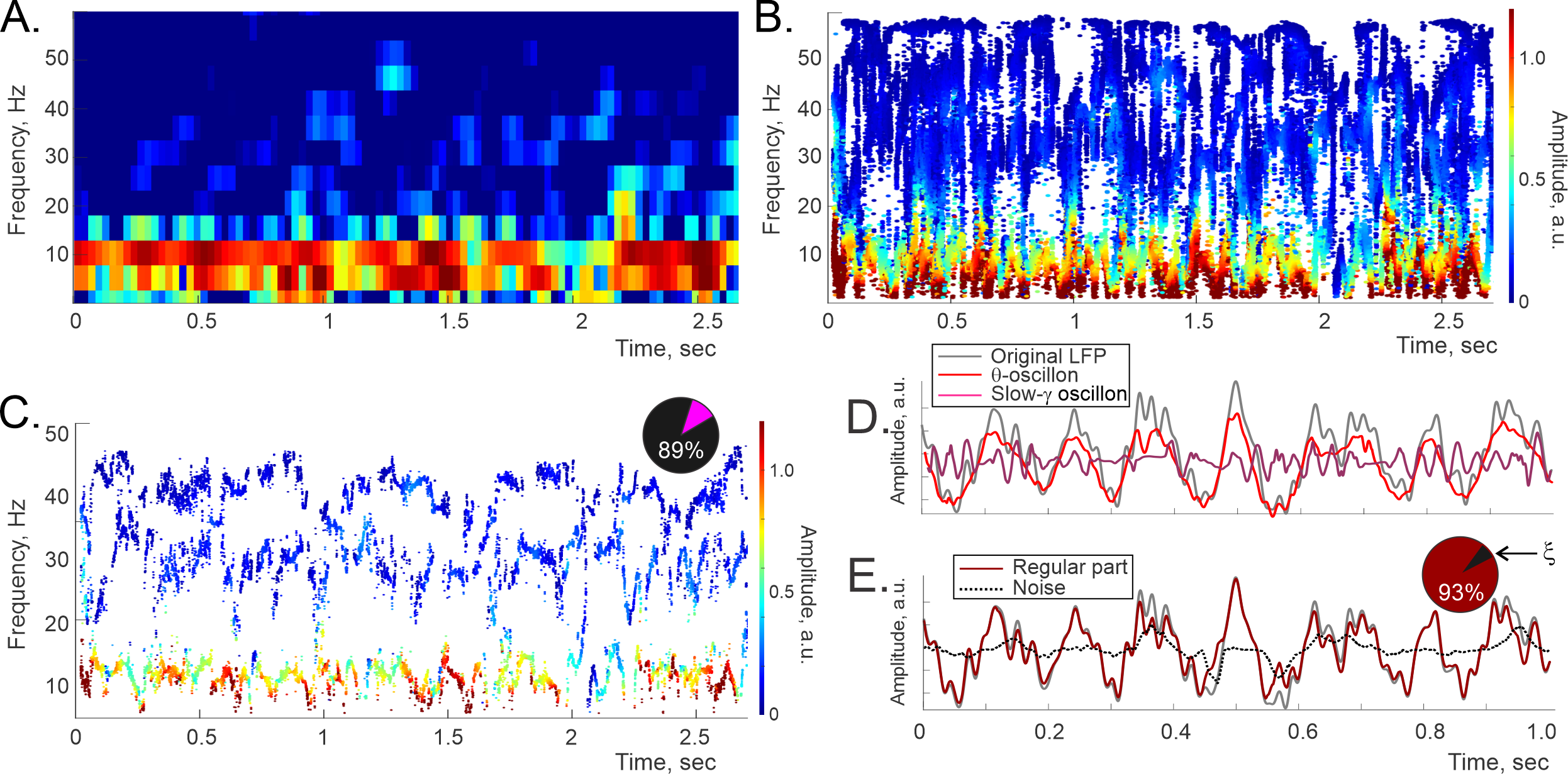}
	\caption{\footnotesize
		\textbf{Oscillons and spectral waves} constructed for the LFP data recorded in the rat's cortex.
		\textbf{A}. Fourier spectrogram: the high-power stripe between about $4$ and $12$ Hz marks the 
		conventional $\theta$-band, the slow-$\gamma$ domain lays approximately between $20$ and $45$ Hz.
		\textbf{B}. The corresponding full Pad\'e spectrogram, same time resolution, shows the pattern of
		``flexible" frequencies, both stable and unstable. The vertical position of each dot marks a specific
		frequency value, the horizontal position marks the corresponding time, and the color indicates the
		instantaneous amplitude (colorbar on the right).
		\textbf{C}. Most frequencies, (typically over $80\%$, dark section on the pie diagram), are unstable
		or ``noise-carrying." Removing them reveals the regular Pad\'e spectrogram, on which the stable 
		frequencies trace out regular timelines---\textit{spectral waves}, $\nu_q(t)$. Color change along
		each spectral wave encodes the corresponding time-dependent amplitude, $A_q(t)$ (Eq.~\ref{osc}).
		\textbf{D}. Combining a particular spectral wave, $\nu_q(t)$, with its amplitude, $A_q(t)$, yields
		an individual oscillon, as indicated by Equation~\ref{osc}). Shown is a one-second-long segment of
		the	cortical $\theta$-oscillon (red trace) and the slow-$\gamma$ oscillon (pink trace). Notice that
		summing	just these two oscillons (first two terms in the Eq.~\ref{oscs}) already approximates the
		full LFP profile (gray line) quite closely. 
		\textbf{E}. The superposition of all the inputs with regular frequencies (dark red trace), closely
		matches	the amplitude original signal (gray trace). The difference is due to the noise component, 
		$\xi(t)$, carried by the unstable frequencies (dotted black line) typically accounts for less than
		$5-7\%$ of the signal's net power during active behaviors (dark section on the pie diagram), and about
		$10-15\%$ during quiescence. Data sampled in a $6$ months old, wake male rat during active behavior.
	} 
	\label{fig:DPT} 
\end{figure}

Synchronized neural activity induces rhythmically oscillating electrical fields that modulate circuit dynamics
at multiple spatiotemporal scales \cite{Buzsaki1,ClgRh,BuzRh}. The corresponding local field potentials (LFPs)
are easily recorded and widely used for describing varieties of neurophysiological phenomena \cite{Thut,Kopell}.
However, our perspective on the LFPs' structure and properties, as well as our interpretation
of its functions depend inherently on the techniques used in data analyses. Currently, most studies are based
on Fourier paradigm, in which the oscillating LFPs are viewed as a superpositions of harmonics with constant
frequencies \cite{Brigham,Vugt}. The familiar $\theta$, $\gamma$, and other ``brain waves" are combinations of
such harmonics, occupying a particular frequency band \cite{ClgRh,BuzRh} (Fig.~\ref{fig:DPT}A).

Despite their commonality, Fourier decompositions are known to provide a limited detalization of LFP dynamics,
particularly for transient, irregular or noisy data, and thus may obscure the brain wave design. A recently 
proposed  alternative, the \textit{Discrete Pad\'e Transform}\footnote{Throughout the text, terminological
	definitions are given in \textit{italics}.} (DPT), allows replacing the packets of Fourier harmonics with
solitary waves that adapt their frequencies to the fields' ongoing structure \cite{Bessis1,Bessis2,Perotti1}
(Fig.~\ref{fig:DPT}B). As it turns out, there are two kinds of such ``flexible" frequencies: those that change
over time in a regular manner, leaving distinct, contiguous traces on the spectrogram, and those that assume
sporadic values from moment to moment. Mathematically, the ``irregular" harmonics represent the noise, $\xi(t)$,
whereas the ``regular" frequencies define the signal's oscillatory part \cite{Oscs1,Zobaer}
(Fig.~\ref{fig:DPT}C, Sec.~\ref{Four}).

At high temporal resolutions,  the ``timelines" of regular frequencies produce undulant patterns---the \textit{
	spectral waves} (Fig.~\ref{fig:DPT}C). These waves vacillate around their respective lento-changing means,
and span over the frequency domains attributed to the traditional, Fourier-defined rhythms. For example, the
lowest-frequency spectral wave, prominent during active behavior, occupies the range that roughly corresponds
to the traditional $\theta$-domain \cite{Jung,BzTh1,BzTh2,BurTh,ColginTh}. The following spectral wave occupies
the slow-$\gamma$ domain \cite{ClgGm}, and so forth (Fig.~\ref{fig:DPT}C). The gap between the mean frequencies
is typically larger than the undulations' magnitudes, which allows keeping the standard nomenclature, e.g.,
using the notation $\nu_\theta(t)$ for the $\theta$-domain spectral wave, $\nu_{\gamma_s}(t)$ for the 
slow-$\gamma$ spectral wave, etc. Each contiguous frequency, $\nu_q(t)$, contributes with a certain amplitude,
$A_q(t)$, and a phase, $\phi_q(t)$, giving rise to a compound wave,
\begin{equation}
	\vartheta_q(t)=A_{q}e^{i\phi_{q}(t)},
	\label{osc}
\end{equation}
which we refer to as \textit{brain wave oscillon}, or just \textit{oscillon} for short \cite{Oscs1,Zobaer} 
(Fig.~\ref{fig:DPT}B, Sec.~\ref{Pade}). The superposition of the oscillons and the noise, $\xi(t)$, amounts to
the original LFP,
\begin{equation}
\ell(t) = A_{\theta} e^{i\phi_{\theta}(t)} +A_{\gammaup_s} e^{i\phi_{\gamma_s}(t)}+\ldots+\xi(t).
\label{oscs}
\end{equation} 
Typically, the main contribution is made by the $\theta$-oscillon, then the next biggest contribution comes
from the slow-$\gamma$ oscillon, etc. (Fig.~\ref{fig:DPT}D).

Importantly, all oscillon parameters are obtained empirically from the recorded data: the amplitudes, the
shapes of the spectral waves and the noise exhibit robust, tractable characteristics that reflect physical
organization of the synchronized neuronal activity \cite{Arenas1,Liao,Mi,Restrepo,Burton}. In other words,
the elements of the decomposition (\ref{oscs}) can be interpreted not only as structural motifs, but also as
functional units within the LFP waves. This raises many questions: what are the physiological properties of
the oscillons? Which of their features were previously captured through Fourier analyses, and how accurately?
Do the oscillons reveal any new operational aspects of the brain waves?

In the following, we focus specifically on the high-amplitude, low-frequency $\theta$-oscillons, recorded in
the hippocampal CA1 area and in the cortex of male rats, shuttling between two food wells on a linear track.
The experimental specifics are briefly outlined in Section~\ref{sec:emt}, (computational techniques and 
mathematical notions are summarily reviewed in the Supplement, Sec.~\ref{sec:met}, for more details see
\cite{Bessis1,Bessis2,Perotti1,Oscs1,Zobaer}). In Section~\ref{sec:res}, we demonstrate that although some
qualities of the $\theta$-oscillons parallel those of the traditional (i.e., Fourier-defined) $\theta$-waves,
many of their characteristics have not been resolved or even addressed by conventional analyses.
In Section~\ref{sec:kur}, we use a basic model of neuronal synchronization to illustrate and contextualize 
our empirical observations. In Section~\ref{sec:dis}, we summarize the results and point out a new vantage
point on the information exchange in the hippocampo-cortical network suggested by the newfound architecture
of the $\theta$-rhythmicity.

\section{Methods}
\label{sec:emt}


\begin{wrapfigure}{c}{0.3\textwidth}
	\includegraphics[scale=0.6]{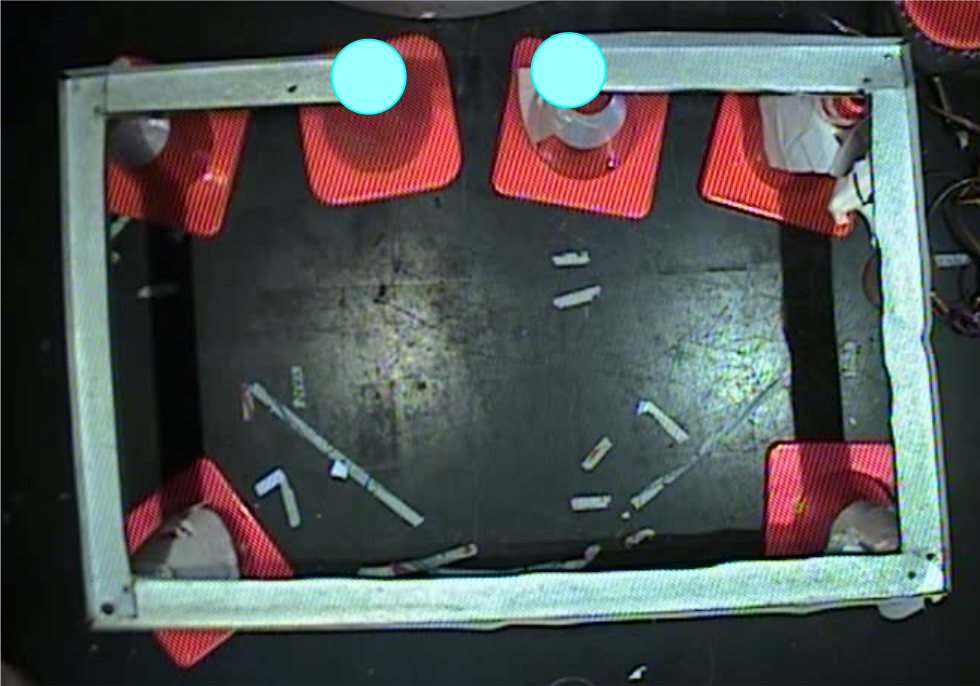}
	\caption{\footnotesize
		\textbf{Linear track} explored by the rat is $3.5$ meters long, a typical run between the food wells
		(blue dots) takes about $30$ sec. Fast moves over the straight segments last $5-6$ secs, slowdowns
		occur at the corners and during feeding.
	}
	\label{fig:track} 
\end{wrapfigure}

\textbf{Experimental procedures}. Data was recorded from the CA1 pyramidal cell layer of Long-Evans rats' 
hippocampus and from anterior cingulate cortex \cite{Carli}. The animals were trained in a familiar
room to run back and forth on a rectangular track for food reward (Fig.~\ref{fig:track}). 
The daily recording procedure consisted of two $30$ min running sessions. Two diodes were mounted over the
animal's head to track its positions, at $33$ Hz rate, with a spatial resolution of about $2$ mm. Further 
details on surgery, tetrode recordings and other procedures can be found in \cite{Carli}. The data analyzed
here are not included in \cite{Carli}, but experimental details are identical.

\textbf{Signal processing}. The original LFP data was sampled at $S_r=8$ kHz rate, and interpolated to $\tilde
{S}_r=36$ kHz rate, to improve the quality of the reconstructions at low frequencies. The signals were filtered
between $0$ and $60$ Hz, and downsampled $2\leq m\leq 4$ times, producing $m$ nearly identical, interlaced 
subseries, which helped to ensure robustness and consistency of the results. These and other procedures
permitted using high time resolutions ($T_W\approx 50-75$ ms time windows), simultaneously with high frequency
resolution of the LFP dynamics (Sec.~\ref{sec:met}).

\section{Results}
\label{sec:res}

The animal's trajectory consisted of intermittent sequence of fast moves along the straight segments of the
track, slowdowns at the turning points and the periods when the animal consumed his food reward 
(Fig.~\ref{fig:track}). 
The following analyses target specifically the $\theta$-oscillons' structure during fast moves.

\textit{Shape of the oscillons}.  Similarly to their Fourier counterparts---the traditional hippocampal 
($f_{\theta}^h$) and cortical ($f_{\theta}^c$) $\theta$-waves---the $\theta$-oscillons, $\vartheta_{\theta}^h$
and $\vartheta_{\theta}^c$, increase their amplitudes during active moves (speed $s$ over $2$ cm/sec) and
diminish (but do not vanish) during slowdowns and quiescence \cite{BzTh1,BzTh2,BurTh,ColginTh}. Due to the
removed noise, the amplitudes of the hippocampal and cortical $\theta$-oscillons' are lower than the amplitudes
of the corresponding Fourier-waves, on  average, by $5-10\%$. The \textit{average} frequency span of the
hippocampal $\theta$-oscillons is between $3$ and $13.5$ Hz (mean over $12$ four-second-long wave segments),
and the cortical $\theta$-oscillons are typically confined between $4$ and $12.5$ Hz. Although the individual
spectral waves' fringes can escape these limits or remain confined within much narrower strips, their spreads
generally conform with the traditional $\theta$-bandwidths (Fig.~\ref{fig:DPT}A,B).

\begin{figure} 
	\includegraphics[scale=0.8]{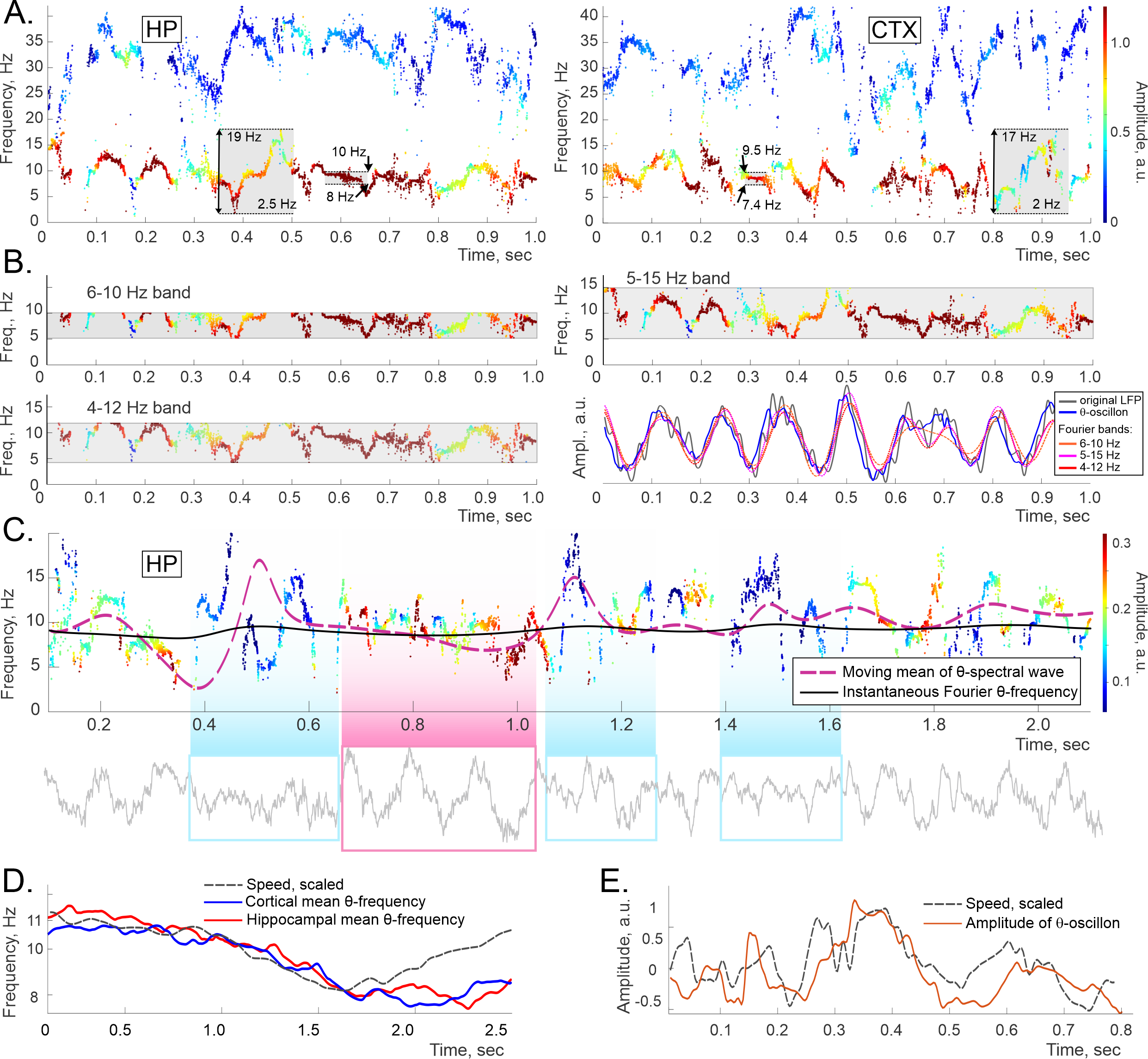}
	\caption{\footnotesize
		\textbf{A}. The lowest spectral wave occupies the domain that is generally attributed to the 
		$\theta$-frequency band. The color of each dot represents the amplitude, as on Fig.~\ref{fig:DPT}.
		The spectral widths range from about $17$ to about $2$ Hz (gray boxes).
		\textbf{B}. The hippocampal $\theta$-oscillon's spectral wave, made visible through three ``frequency
		slits" that represent three most commonly used $\theta$-bands, $4$-$12$ Hz, $5$-$15$ Hz, and $6$-$11$
		Hz (gray stripes). The frequencies that fit into a slit are the ones that produce the corresponding
		Fourier wave, shown	as red-shaded traces on the bottom right panel. Note that the spectral waves
		are crosscut by all $\theta$-bands. The Fourier waves are close	to each other (DTW distances 
		$D(f_{\theta_1}^h,f_{\theta_2}^h)\approx6\%$, $D(f_{\theta_2}^h,f_{\theta_3}^h)\approx 5\%$,
		$D(f_{\theta_3}^h,f_{\theta_1}^h)\approx 8\%$), but further from the $\theta$-oscillon (e.g., 
		$D(\vartheta_{\theta}^h,f_{\theta_2}^h)\approx 9\%$), which, in turn, is closer to the original
		LFP (the $\ell(t)$ in Eq.~\ref{oscs}, gray trace, $D(\ell^h,\vartheta_{\theta}^h)=4\%$).
		\textbf{C}. A longer segment of a hippocampal-$\theta$ spectral wave. The nearly-constant solid black
		trace in the middle shows the instantaneous Fourier $\theta$-frequency. The dashed purple line shows
		the spectral wave's moving average, which clearly	provides a more detailed description of the
		$\theta$-rhythm's trend. The waveform (bottom panel) is regular when the mean is steady (pink box),
		and corrugates when the mean is perturbed (blue boxes).
		\textbf{D}. The moving mean of the hippocampal and cortical $\theta$-oscillons' spectral waves (blue
		and red curves respectively) follow the speed profile (dashed gray curve). The latter was scaled 
		vertically to match, for illustrative purposes, see also Fig.~\ref{fig:sp}.
		\textbf{E}. The  amplitude of the hippocampal $\theta$-oscillon also co-varies with the speed, which
		was captured previously via Fourier analyses \cite{Young,ChenZ,Kennedy}.
	}
	\label{fig:mean} 
\end{figure}

To put these observations into a perspective, note that there exist several consensual $\theta$-frequency bands,
e.g., from $4$ to $12$ Hz, from $5$ to $10$ Hz, from $6$ to $10$ Hz, from $6$ to $12$ Hz, from $5$ to $15$ Hz,
etc. \cite{BranTh,HuxTh,MizusTh,HarTh,Jezek,BrunG,KropffTh,Richard,Young,ChenZ,Ahmed,Zheng,Kennedy}. The shapes
of the corresponding Fourier $\theta$-waves differ by about $\sim 6\%$, as measured by the relative amount of 
local adjustments required to match a pair of waveforms (the Dynamic Time Warping method, DTW \cite{Brndt,Salva,
Neam}, see Fig.~\ref{fig:dtw}). In other words, since bandwidths are obtained through empirically established
frequency limits, the resulting wave shapes are fuzzily defined. In contrast, the frequency profiles of the
$\theta$-oscillons are determined squarely, over each time interval, by the shapes of their respective spectral
waves, which may, at times, drop below $3$ Hz, get as high as $20$ Hz, or narrow into a tight $1-2$ Hz band that
typically stretches along the mean $\theta$-frequency (Fig.~\ref{fig:mean}A,B). 

Overall, the shapes of $\theta$-oscillons are similar to the shapes of the Fourier-defined $\theta$-waves: the
DTW-difference between them during active moves are small, $D(\vartheta_{\theta}^h,f_{\theta}^h)\approx 6.4\pm
1.2\%$ and $D(\vartheta_{\theta}^c,f_{\theta}^c) \approx7.1\pm1.5\%$ respectively, relative to the waveforms'
length. However, adaptively built $\theta$-oscillons capture the ongoing details of the LFP waves better and
are hence DTW-closer to the original LFP signal ($D(f_{\theta}^h,\ell^h)\approx(4\pm0.7)D(\vartheta_{\theta}^h,
\ell^h)$, $D(f_{\theta}^c,\ell^c)\approx(5\pm0.9)D(\vartheta_{\theta}^c,\ell^c)$, values computed for $17$
waveforms, for an illustration see Fig.~\ref{fig:mean}B and Fig.~\ref{fig:thgm}). Between themselves, the shapes
of the hippocampal and the cortical oscillons differ about as much as the traditional, Fourier, $\theta$-waves:
over $1-2$ secs segments, $D(\vartheta_{\theta}^h,\vartheta_{\theta}^c)\approx D(f_{\theta}^h,f_{\theta}^c)
\approx7\pm 0.8\%$.

\textit{Coupling to locomotion}. The $\theta$-oscillons' amplitudes and their mean frequencies are coupled to
speed, as one would expect based on a host of previous studies of $\theta$-rhythmicity \cite{KropffTh,Richard,
	Young,Kennedy}. This correspondence is nevertheless informative since oscillons are qualitatively different
objects. While the traditional brain waves' frequency is evaluated by tracking their Fourier-envelope 
(Fig.\ref{fig:hil}), the oscillons' mean frequency is obtained as their spectral waves' moving mean. 
As illustrated on Figs.~\ref{fig:mean}C, the two outcomes are consistent, but the latter provides a more nuanced
description of the oscillatory trends: both hippocampal and the cortical $\theta$-oscillons' mean frequencies,
$\nu_{\theta,0}^h$ and $\nu_{\theta,0}^c$, vary more intricately than their Fourier counterparts, which shows
only limited, sluggish changes. Correspondingly, the co-variance of the oscillons' mean frequency with the rat's
speed is much more salient (Figs.~\ref{fig:mean}D). Similarities between the two descriptions can be made more
conspicuous by scaling the Fourier-frequency up by an order of magnitude (Fig.~\ref{fig:sp}A), which further
illustrates that the traditional descriptions capture the mean $\theta$-oscillon dynamics at a diminished scale.
Similar mean behavior is exhibited by the slow-$\gamma$ oscillon also co-varies with speed (Fig.~\ref{fig:sp}B).

\textit{Spectral undulations} are the oscillons' distinguishing feature that capture detailed aspects of the
$\theta$-rhythm's cadence (Fig.~\ref{fig:mean}C). The semi-periodic appearance of the spectral waves suggests
that their ups and downs should be decomposable into a harmonics series,
\begin{equation}
	\nu_\theta(t)=\nu_{\theta,0} + \nu_{\theta,1} \cos(\Omega_{\theta,1}t +
	\phi_{\theta,1}) +\nu_{\theta,2}\cos(\Omega_{\theta,2}t +\phi_{\theta,2}) +\ldots\, ,
	\label{emb}
\end{equation}
where the modulating, embedded frequencies\footnote{The suppression of the ``$h$" and ``$c$" superscripts 
here and below indicates that a formula or a notation applies to both the hippocampal and the cortical cases.},
$\Omega_{\theta,i}$, the corresponding modulation depths, $\nu_{\theta,i}$s, and the phases, $\phi_{\theta,i}$
evolve about as slowly as the mean frequency, $\nu_{\theta,0}$ (Fig.~\ref{fig:mean}D). By itself, this 
assumption is patently generic---given a sufficient number of terms, suitable expansions (\ref{emb}) can be
produced, over appropriate time periods, for almost any data series \cite{Corduneanu}. Question is, whether
spectral waves' expansions can be succinct, hold over sufficiently long periods and whether their terms are
interpretive in the context of underlying physiological processes.

\begin{figure} 
	\includegraphics[scale=0.75]{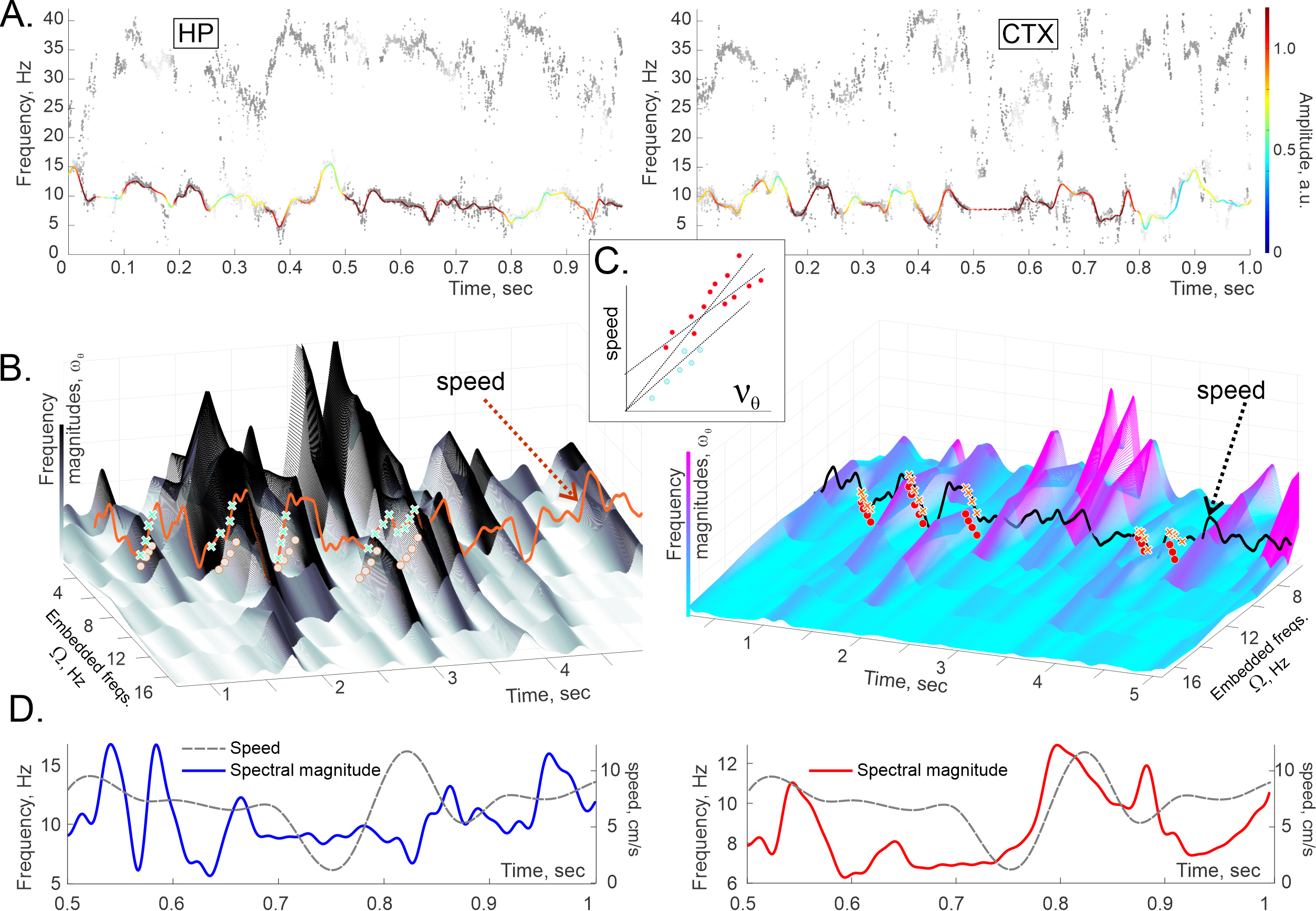}
	\caption{\footnotesize
		\textbf{Spectral waves and embedded frequencies}. 
		\textbf{A}. The spectral patterns produced via shifting-window evaluation of instantaneous frequencies
		are intermittent (Fig.~\ref{fig:mean}A). To recapture the underlying continuous spectral dynamics, we
		interpolated the raw datapoints over a uniform time series, thus recovering the hippocampal (left) and
		the cortical (right) spectral waves with uninterrupted shapes.
		\textbf{B}. The contiguous data series allow constructing $3D$ Welch's spectrogram on which each peak
		along the frequency axis highlights a particular embedded frequency. Altitudinal shadowing emphasizes
		higher peaks (colorbar along the vertical axis). Note that most peaks in both hippocampal (left) and
		the cortical (right) $W$-spectrograms are localized not only in frequency but also in time, indicating
		short-lived perturbations of spectra. The dynamics of these frequencies is coupled with the 
		speed---higher speeds drive up the magnitudes of the embedded frequencies. The speed profile is scaled
		vertically and shifted horizontally to best match the frequency magnitudes (orange and black trace
		respectively). While the response of the hippocampal frequency to speed is nearly immediate (about
		$\tau=70-120$ ms delay), the cortical response is delayed by about $\tau=300\pm50$ ms.
		\textbf{C}. Examples of the individual cortical peaks' magnitudes, sampled at random, over particular
		frequencies	(heights of the dots on the panels B) and the corresponding speeds (heights of the crosses)
		exhibit	clear quasi-linear dependencies. The exponents of these dependencies may slightly change from
		one	appearance of a peak (an instantiation of an embedded frequency) to another.
		\textbf{D}. The net magnitude of the spectral wave ($\theta$-bandwidth) co-varies with the speed with
		the	delay of $\tau=100$ ms, depending on the case, in the hippocampus (left panel) and an extra
		$\tau=300$ ms in the cortex (right panel).
	} 
	\label{fig:spw} 
\end{figure}

To test these possibilities, we interpolated the ``raw" frequency patterns over uniformly spaced time points
and obtained contiguous spectral waves over about $5.5$ secs---about $1/6$ of a typical trip between the food
wells (Fig.~\ref{fig:spw}A). The full wave was then split into $600$ ms long, strongly overlapping (by $99.9\%$)
segments, for which we computed power profiles using Welch's method \cite{Welch,Proak}, see Sec.~\ref{sec:met}.
Arranging the results along the discrete time axis yields three-dimensional ($3D$) Welch's spectrograms 
(\textit{$W$-spectrograms}), whose lateral sections consist of the familiar instantaneous power-frequency
profiles, with peaks marking the embedded frequencies, and whose longitudinal sections show the time dynamics
of these peaks (Fig.~\ref{fig:spw}B, see also Fig.~\ref{fig:w} in Sec.~\ref{Data}).

The results demonstrate that both hippocampal and cortical spectral waves have complex dynamics. First, most
peaks are localized not only in frequency but also in time: a typical peak grows and wanes off in about $200-
300$ ms, i.e., the embedded frequencies and the depth of modulation are highly transitive, changing faster than
the mean frequencies by an order of magnitude \cite{Zobaer}. Thus, the representation (\ref{emb}) holds over 
relatively short periods (typically $1$ sec or less, see \cite{Oscs1}), and then requires corrections in order
to account for rapidly accumulating changes. In other words, the $\theta$-oscillons may be viewed as relatively
steady oscillatory processes, altering at the behavioral timescale at around $\sim\nu_{\theta,0}\approx 8$ Hz,
and modulated by series of swift, transient vibrations. 

Curiously, certain peaks in the $W$-spectrograms appear and disappear repeatedly near the same location along
the $\Omega$-axis, i.e., fast moves can repeatedly incite $\theta$-rhythm's vibrations at the same embedded
frequencies (Fig.~\ref{fig:spw}B, Fig.~\ref{fig:w5}), indicating restorative network dynamics \cite{Go}.
Other peaks appear sporadically, possibly reflecting spontaneously generated oscillations, resonances or brief
external contributions (Fig.~\ref{fig:spw}B, \cite{Zobaer}). In either case, these events are coupled with the
rat's ongoing behavior: power flows into higher embedded frequencies ($\Omega_i$ over $4-5$ Hz) as the speed 
increases, and recedes as the speed drops, i.e., fast moves drive spectral undulations both in the hippocampus
and in the cortex. Markedly, cortical responses tend to delay by about $\bar{\tau}^c=200-300$ ms, while the
mean hippocampal delay is shorter, about $\bar{\tau}^h=70-120$ ms (Fig.~\ref{fig:spw}B). With these shifts 
taken into account, spectral waves are structurally closer to one another than the corresponding oscillons
$D(\nu_{\theta}^h,\nu_{\theta}^c)\approx (0.8\pm 0.12) D(\vartheta_{\theta}^h,\vartheta_{\theta}^c)$, which
may be viewed as a manifestation of the hippocampo-cortical frequency coupling at the circuit level. 

The peak magnitudes associated with the individual embedded spectral frequencies $\Omega_i$ (the coefficients
of the oscillatory terms in the Equation~\ref{oscs}) tend to grow roughly proportionally to speed, 
\begin{equation}
	\nu_{\theta,i}(t)\approx \alpha_{i}+w_{i}\cdot s^{\kappa_i}(t-\tau_{i}),
	\label{oscmod}
\end{equation}
where the hippocampal and cortical exponents are close to $1$ ($\bar{\kappa}^h=1.2\pm0.44$ vs. $\bar{\kappa}^c
= 0.9\pm0.35$, evaluated for $10$ peaks each). The coefficients $\alpha_{i}$, $w_i$ and the temporal lags of
the peaks' responses, $\tau_i$, vary by about $10-25\%$ from one embedded frequency to another, in both brain
areas. 

The coupling (\ref{oscmod}) between the embedded frequency magnitudes and speed is foreseeable: linear 
modulation of extracellular fields' base frequency by speed,
\begin{equation}
	\nu_{\theta} = \nu_{\theta,0}+\beta\cdot s(t)\cdot\cos(\phi(t)),
	\label{kif}
\end{equation}
was hypothesized by J. O'Keefe and M. Recce $30$ years ago, upon discovery of coupling between $\theta$-phase
shift, $\phi(t)$, and spiking probability \cite{Recce}. This dependence was thenceforth used in oscillatory
interference models for explaining the hippocampal place cell \cite{Recce,OscG} and the entorhinal grid cell
\cite{OscG1,OscG2,OscG3} firing. Several \textit{in vitro} experiments \cite{Giocomo1,Giocomo2,Shay} and 
Fourier-based LFP analyses \cite{Jeewajee} provide supporting evidence for these models (see however 
\cite{Domnisoru}). Here we find that $\theta$-oscillons' dynamics, obtained directly from \textit{in vivo}
electrophysiological data conform with the previous studies. However, the hypothesized proportionality between
the $\theta$-frequency and speed is approximate---the actual dependence (\ref{oscmod}) may deviate from strict
linearity. Second, the frequency expansion (\ref{emb}) may contain several oscillatory, speed-modulated terms,
whereas the model expansion (\ref{kif}) contains only one (Fig.~\ref{fig:sol}). Third, the modulation parameters
are particularized, i.e., peak- and time-specific, albeit distributed around well-defined means. 
Thus, the $\theta$-oscillons' structure fleshes out many details and exhibits the actual, empirical properties
of speed-controlled oscillators driving the spectral wave (Fig.~\ref{fig:spw}B,C,D). Note here that applying 
$W$-spectrogram analyses to the instantaneous Fourier frequencies do not capture any of these structures---the
detailed spectral dynamics is averaged out (Fig.~\ref{fig:fw}).

\textit{Noise}. The qualitative difference between the regular and the irregular frequencies allows delineating
the LFP's noise component. While in most empirical studies ``noise'' is identified \textit{ad hoc}, as a
cumulation of irregular fluctuations or unpredictable interferences within the signal  \cite{Faisal,Ermentrout,
Stein,Rowe,McDonnell}, here noise is defined conceptually, based on intrinsic properties of Pad\'e approximants
to the signal's $z$-transform that are mathematically tied to stochasticity (Sec.~\ref{sec:met}, \cite{Bessis1,
	Bessis2,Perotti1}). Being that noise is qualitatively distinct from the oscillatory part, its dynamics 
provides complementary description of the network's state.

\begin{figure} 
	\includegraphics[scale=0.8]{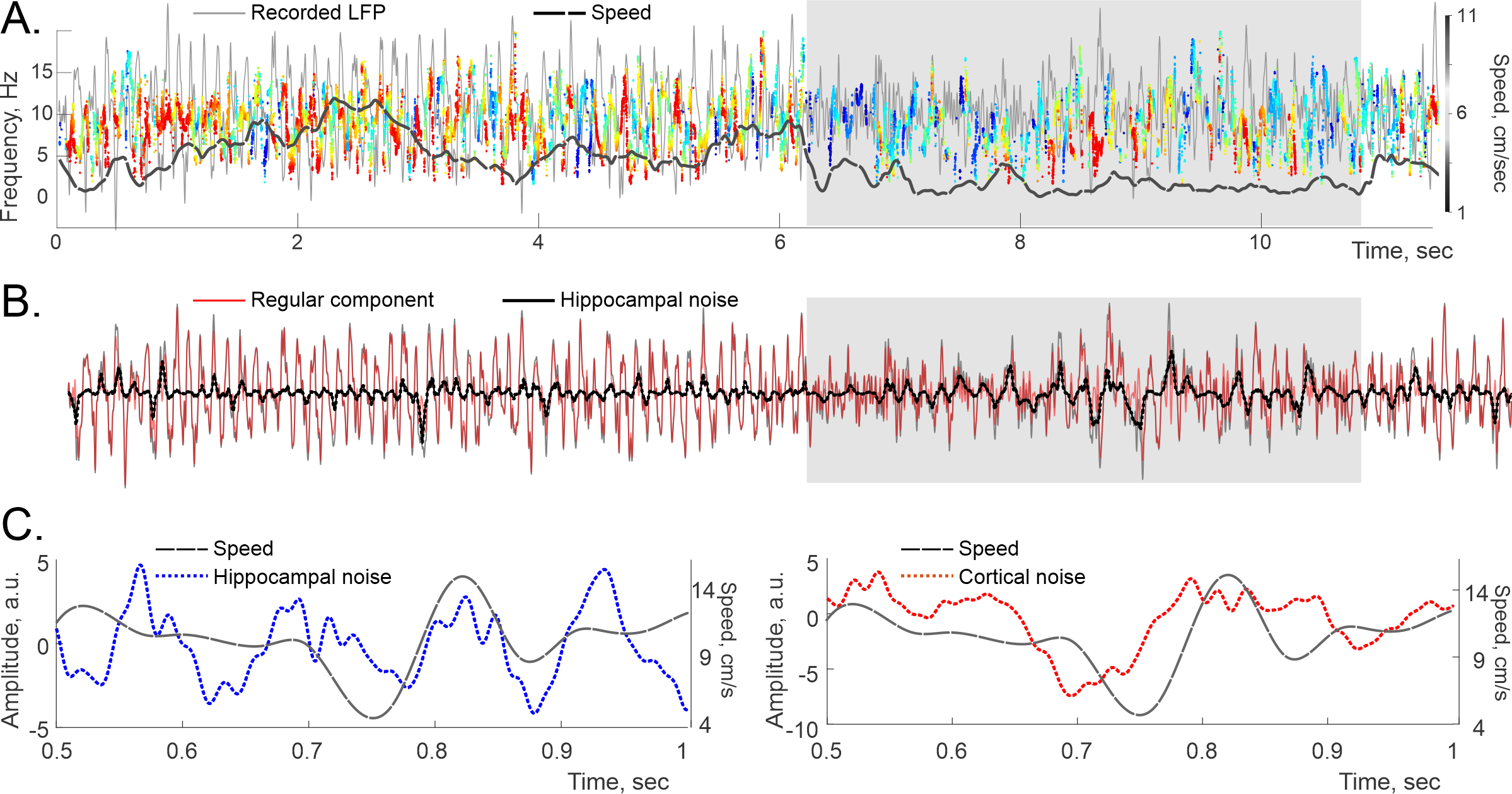}
	\caption{\footnotesize \textbf{Spectral wave magnitude, noise and speed}.
		\textbf{A}. A segment of a hippocampal spectral wave shows magnitude increase during activity
		with speed, which reflects increased level of synchronization (see below). Shaded area highlights a 
		period of slow motion, during which the noise escalates. The original LFP amplitude is shown by the gray
		trace in the background, for reference. The dashed black curve shows the animal's speed. 
		\textbf{B}. The dynamics of the regular part of the LFP (red trace) and the noise component (dotted black
		trace), obtained for a $12$-second lap. Original LFP is shown by gray trace in the background.
		\textbf{C}. The hippocampal (left panel) and cortical (right panel) noise levels follow speed, but more
		loosely than the oscillon's amplitude.
	}
	\label{fig:noi}
\end{figure}

As indicated on Fig.~\ref{fig:DPT}C, in a typical LFP, only a few percent of the frequencies exhibit regular
behavior, and yet their combined contribution is dominant: the stochastic component, $\xi(t)$, usually accounts
for less than $5\%$ of the signal’s amplitude, i.e., the noise level is generally low \cite{Oscs1,Zobaer}. 
Structurally, the DPT noise component has undulatory appearance and can make positive or negative contributions
to the net signal, modulated by the rat's physiological state \cite{Perotti2,Zobaer,Oscs1}, Fig.~\ref{fig:noi}A,
B. First, the noise amplitude tends to grow with speed (Fig.~\ref{fig:noi}C), which suggests that increase of
stochasticity is associated with the surge of the modulating vibrations (Fig.~\ref{fig:spw}). However, the speed
couples to noise weaker than to the oscillatory $\theta$-amplitude or to the spectral wave (cortical separation
$D(\xi^c,s)=12.2\pm2\%$, hippocampal $D(\xi^h,s)=10.6\pm1.5\%$, Fig.~\ref{fig:spw}D). Yet, both hippocampal and
cortical noise levels exhibit affinity, $D(\xi^h,\xi^c)\approx14 \pm 2\%$. Most strongly, the noise is coupled
to the amplitude of the oscillon that it envelopes, $D(\vartheta_{\theta}^h,\xi^h)=1.2\pm0.5\%$, $D(\vartheta_{
	\theta}^c,\xi^c)=1.5\pm 0.5\%$, i.e., noise grows and subdues with the ups and downs of the physical
amplitude of the extracellular field (Fig.~\ref{fig:noi}B). As the animal moves out of the $\theta$-state, noise
amplifies by an order of magnitude and decouples from the locomotion, indicating an onset of a ``non-theta" state
\cite{Mysin}, in which the amplitude of $\theta$-oscillon drops by about $50-80\%$ (Fig.~\ref{fig:noi}B, see also
\cite{Hoffman}).

\section{Kuramoto oscillon}
\label{sec:kur}

What is the genesis of the oscillons? It is commonly believed that the rhythmic LFP oscillations emerge from
spontaneous synchronization of neuronal activity, although the specifics of these processes remain unknown
\cite{Arenas1,Liao,Mi,Restrepo,Burton}. A comprehensive analysis of the recorded data or modeling at the
physiological level of detail is prohibitively complex and technically out of reach. Nevertheless, the essence
of synchronization can be illustrated with simple computational models, which helps clarifying the observed
phenomena. Specifically, the celebrated Kuramoto model allows tracing the onset of synchronization in a
population of oscillating units, using a single parameter that describes the coupling strength \cite{Kuramoto,
	Strogatz}.  Under fairly generic assumptions, these oscillators, or \textit{phasors}, can qualitatively 
represent (map to) the recurring activity of individual neurons \cite{IzhPh2,IzhPh3,HopPh0,HopPh1,HopPh2}, so
that their net population dynamics captures the ebb and flow of the mean extracellular field.

\begin{figure} 
	\includegraphics[scale=0.75]{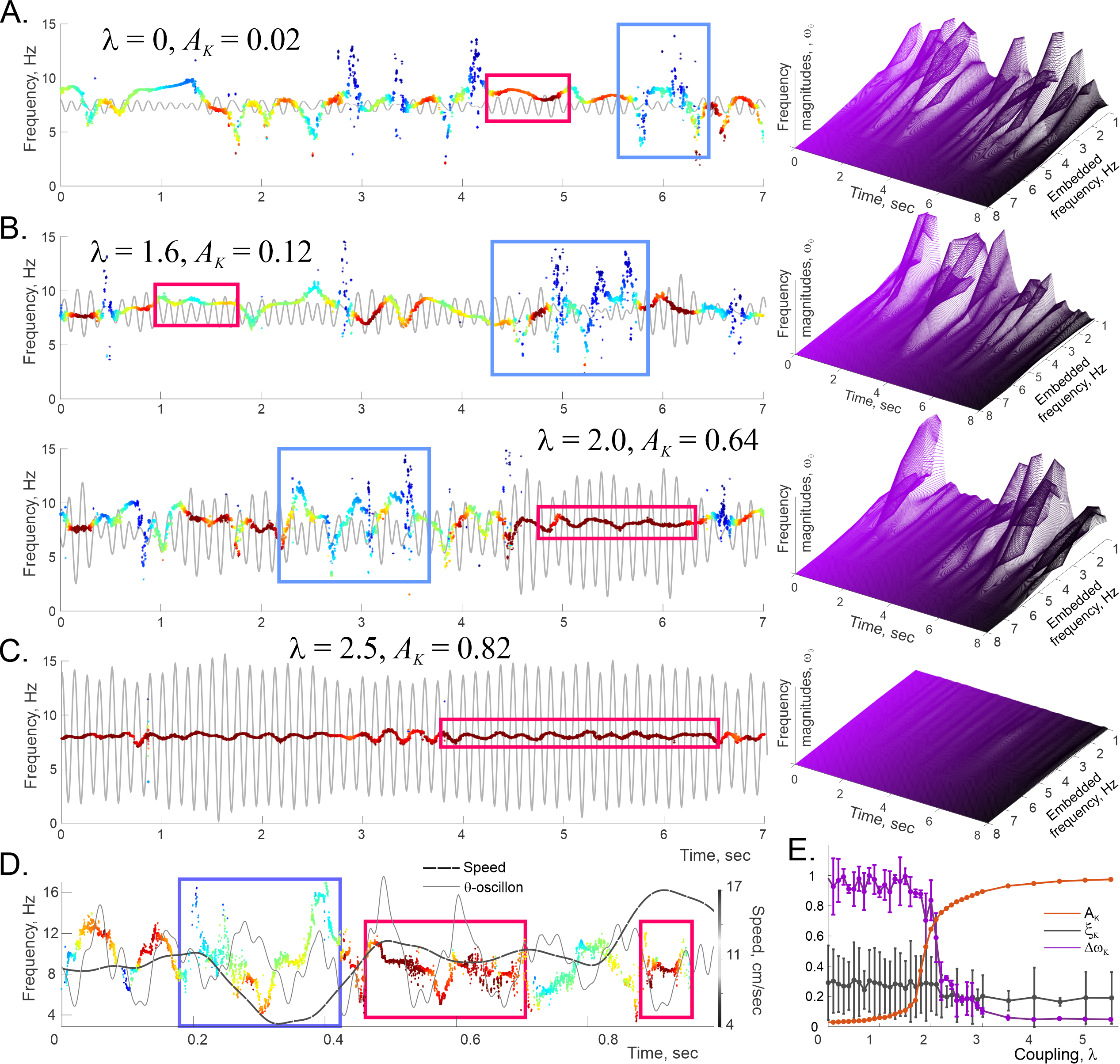}
	\caption{\footnotesize{\textbf{Kuramoto model}. $1000$ oscillators (phasors) with base frequencies normally
		distributed around $8$ Hz with the variance $1$ Hz, coupled via equation~\ref{dyn}, produce a mean field
		characterized by a single spectral wave---a solitary Kuramoto oscillon (gray trace in the background, 
		scaled up on the top panel $10$ times for visibility). On all panels, the maximal amplitude defines the
		color scale. 
		\textbf{A}. At small couplings, $K$-oscillon has low amplitude ($A_K\lesssim 0.02A_{\max}$) and often
		reshapes (the spectral wave then disrupts, blue boxes). The $W$-spectrogram (right panel) shows that
		the embedded frequencies restructure at $\sim 100$ ms timescale.
		\textbf{B}. As the coupling between phasors grows, the synchronized amplitude builds up and the
		$K$-oscillon's shape regularizes. Note that when the spectral wave flattens out, the oscillon is nearly
		sinusoidal (strong synchronization, brown boxes), and the dynamics of the embedded frequencies during
		these periods are suppressed (right panel).
		\textbf{C}. At large couplings, synchronization becomes persistent: the spectral wave narrows, the
		embedded frequencies die out and the oscillon reduces to a nearly-sinusoidal harmonic.
		\textbf{D}. A hippocampal $\theta$-oscillon regularizes its spectral wave and yields higher amplitude
		when the rat's speed is steady (gray dashed line, shifted by $~80$ ms); desynchronization occurs when
		the speed is low or transient.
		\textbf{E}. The oscillon's amplitude, $A_K$ (orange curve), the magnitude of its spectral wave
		(purple curve), and the noise level, $\xi$ (gray curve), for different coupling strengths. As the
		system synchronizes ($1.7\lessapprox\lambda\lessapprox3$), the amplitude grows, while the spectral
		undulations	and the noise subdue. At higher couplings noise is suppressed and
		regular wave dominates. 
	}}
	\label{fig:kur}
\end{figure}

The behavior of each unit is described by a time-dependent phase, $\varphi_m(t)$, that changes between $0$ and
$2\pi$, inducing an oscillatory output, $\ell_m=a_m\,e^{i\varphi_m}$. The field produced by the population of
$M$ phasors is hence
\begin{equation}
	\sum_{m=1}^{M}a_m e^{i\varphi_m(t)}= A_K(t)e^{i\phi_K(t)},
	\label{AK}
\end{equation}
where $A_K(t)$ is the total amplitude and $\phi_K(t)$ the net phase. Each individual phase is coupled to the
mean phase through an equation
\begin{equation}
	\dot{\varphi}_m=2\pi\nu_m+\lambda A_K\sin(\phi_K -\varphi_m),
	\label{dyn}
\end{equation}
where the dot denotes the time derivative and the constant $\lambda$ controls the coupling. More extended
discussions can be found in \cite{Kuramoto,Strogatz}, but in brief, such system can transition between two
phases of behavior. For small couplings, different phases evolve nearly independently, proportionally to
their proper angular speeds, $\varphi_m\approx2\pi\nu_m t$. If these speeds are sufficiently dispersed, the
phases remain out of sync and the net field remains small and irregular, $A_K(t)\approx 0$. As the coupling
increases, the phases become more correlated. If the underlying frequencies are distributed close to a certain
ensemble mean, $\bar{\omega}_m=\omega$, then the synchronized amplitude grows and regularizes with rising
$\lambda$, and eventually produce a synchronous beat with the frequency $\omega$.

From perspective of this discussion, this phenomenon is of interest because it may yield a solitary ``synthetic"
oscillon, that helps illustrating properties of the physiological oscillons. As shown on Fig.~\ref{fig:kur}A,
if the phasors' proper frequencies are distributed closely (within about $\pm 3$ Hz) to the mean frequency of
$\nu_{\theta,0}\approx 8$ Hz, then the net field's dynamics is characterized by a single spectral wave that
changes its properties according to the synchronization level. 
For weak couplings (small $\lambda$s), synchronization is fragmentary: segments with steady frequency extend
over a few oscillations, outside of which the spectral wave has a large magnitude (note that the oscillon's
amplitude, $A_K$, remains low) and is carried by many embedded frequencies, as indicated by abundance of 
transient peaks on $W$-spectrograms (Fig.~\ref{fig:kur}A, right panel). As $\lambda$ grows, the net amplitude,
$A_K$, increases and the segments of synchronicity lengthen, while the embedded oscillations subdue
(Fig.~\ref{fig:kur}B). 
As $\lambda$ grows further, the embedded frequencies reduce in number and loose magnitudes, notably over the
periods of increased synchronicity. As $\lambda$ gets even higher, the oscillon turns into a simple harmonic,
and its spectral wave degenerates into a line (Fig.~\ref{fig:kur}C). Ultimately, spectral undulations get 
suppressed, as synchronization becomes fully dominant. 

These observations suggest that the spectral wave's width (frequency deviation from the mean, $\Delta_{\theta}
(t)=\nu_{\theta}(t)-\nu_{\theta,0}(t)$), may serve as an indicator of the ongoing synchronization level.
From this perspective, the fact that both hippocampal and cortical $\theta$-spectral waves are generally wide
($\pm4$ Hz, see Figs.~\ref{fig:mean}A,C, \ref{fig:spw}A), implies that the physiological synchronization level
is fairly low.

Note that Kuramoto oscillons produced by weakly-coupled ensembles often exhibit brief periods of regularity,
with reduced spectral undulations, usually accompanied by higher amplitudes (Fig.~\ref{fig:kur}A,B). Similar
effects are observed in the empirical, physiological oscillons, where higher-synchronicity episodes, lasting
about $50-70$ ms in cortex and $100-200$ ms in the hippocampus favor steady moves. Conversely, intensification
of spectral undulations---local desynchronization---tends to co-occur with slowdowns and speed-ups 
(Fig.~\ref{fig:kur}D, see also \cite{Hoffman}). Thus, the physiological oscillons' response to higher speed
effectively corresponds to the increase of coupling $\lambda$.

Surprisingly, the Kuramoto model, being fully deterministic, also produces a noise component. At weak couplings,
about $10\%$ of the recovered frequencies are stable, while the other $90\%$ exhibit erratic behavior and 
represent the \textit{emerging} noise, $\xi_K$, which, however, is small and weakens with growing $\lambda$.
Importantly, the distinction between noise and regularity is robust: injecting artificial noise, $\hat{\xi}$,
white or colored, into the Kuramoto field, up to $\hat{\xi}\approx10\xi_K$, does not alter the amplitude of the
de-noised, regular Kuramoto oscillon. As shown on Fig.~\ref{fig:kur}E, in the desynchronized state ($\lambda
\lesssim 2$), the noise accounts for about $17\%$ of the field, and then nearly disappears as the system 
synchronizes. In other words, high noise level may also be viewed as a signature of desynchronization, whereas
regular oscillations dominate in synchronized states. This behavior is also in agreement with the physiological
dynamics: as shown on Fig.~\ref{fig:noi}, the noise level is lower during active moves, and heightened during
quiescence, indicating that synchronization of the hippocampo-cortical network increases during the animal's
activity \cite{Hoffman}.

\section{Discussion}
\label{sec:dis} 

The intricate structure of the synchronized extracellular fields can be anatomized using different decomposition
techniques. The constituents brought forth by a particular decomposition provide a specific semantics for 
reasoning about the LFP functions. Being that these semantics may differ substantially, one may inquire which
approach better reflects the physical structure of the brain rhythms. The oscillatory nature of LFPs suggests
partitioning the signal into Fourier harmonics---an approach that dominates the field since the discovery of
the brain waves \cite{Buzsaki1,ClgRh,BuzRh}. However, its is also known that the Fourier techniques obscure the
structure of noisy and nonstationary data---precisely the kind of signals that are relevant in biology 
\cite{Grunbaum}. In particular, these considerations apply to the LFPs, since their constituents---the
oscillons---have transient structures enveloped by noise, i.e., are by nature noisy and nonstationary 
\cite{Oscs1,Zobaer}.

Since the oscillons are constructed empirically, using a high time-frequency resolution technique, and exhibit
stable, reproducible features that dovetail with theoretical models of synchronization, they likely capture the
physical architecture of the extracellular fields, whereas the traditional, Fourier-defined brain waves provide
approximative descriptions. Thus far, oscillons were observed in rodents' hippocampal and the cortical LFPs, 
but similar structure should be expected in generic brain rhythms. Their systematic analyses should help linking
electrophysiological data to the synchronization mechanisms and reveal the dynamics of the noise component.

Lastly, the oscillons suggest a fresh vantage point for the principles of information transfer in the 
hippocampo-cortical network. Traditionally, the coupling of LFP rhythms to neuronal activity is traced through
modulations of the brain waves' amplitudes and phases \cite{KropffTh,Richard,Young,ChenZ,Ahmed,Zheng,Kennedy}. In
contrast, frequency-modulated (FM) oscillons imply a complementary format, in which a slow-changing mean
frequency defines the ``channel of communication," over which the information is carried by the rapid phase and
frequency alterations, reflecting fast endogenous dynamics and abrupt external inputs \cite{Zobaer,Ziemer}. In
other words, Fourier analyses emphasize amplitude modulation (AM), while the DPT decomposition highlights the
FM principles of information transfer, carried over several discrete channels \cite{IzhPh2,IzhPh3,HopPh0,HopPh1,
	HopPh2}. In the case of $\theta$-rhythms, the AM-format is manifested in familiar couplings of speed with
the slowly changing mean frequencies, amplitudes, narrowing and widening of the $\theta$-band, etc., whereas
information about rapid activities appears to transmit across the hippocampo-cortical network via alterations
of the embedded $\theta$-frequencies.

\vspace{10pt}
\textbf{Acknowledgments}. The work was supported by NIH grants R01NS110806 (MZ and YD), R01AG074226 (NL and YD),
R01DA054977 (CD and DJ) and NSF grant 1422438  (CH and YD). 

\newpage
\section{Appendix}
\label{sec:sup}
\subsection{Supplementary figures}


\renewcommand{\figurename}{Suppl. Fig.}
\setcounter{figure}{0}
\renewcommand{\thefigure}{S\arabic{figure}}
\renewcommand{\theHfigure}{S\arabic{figure}}


\renewcommand{\figurename}{Fig.}
\begin{figure}[h]
	\centering
	\includegraphics[scale=0.79]{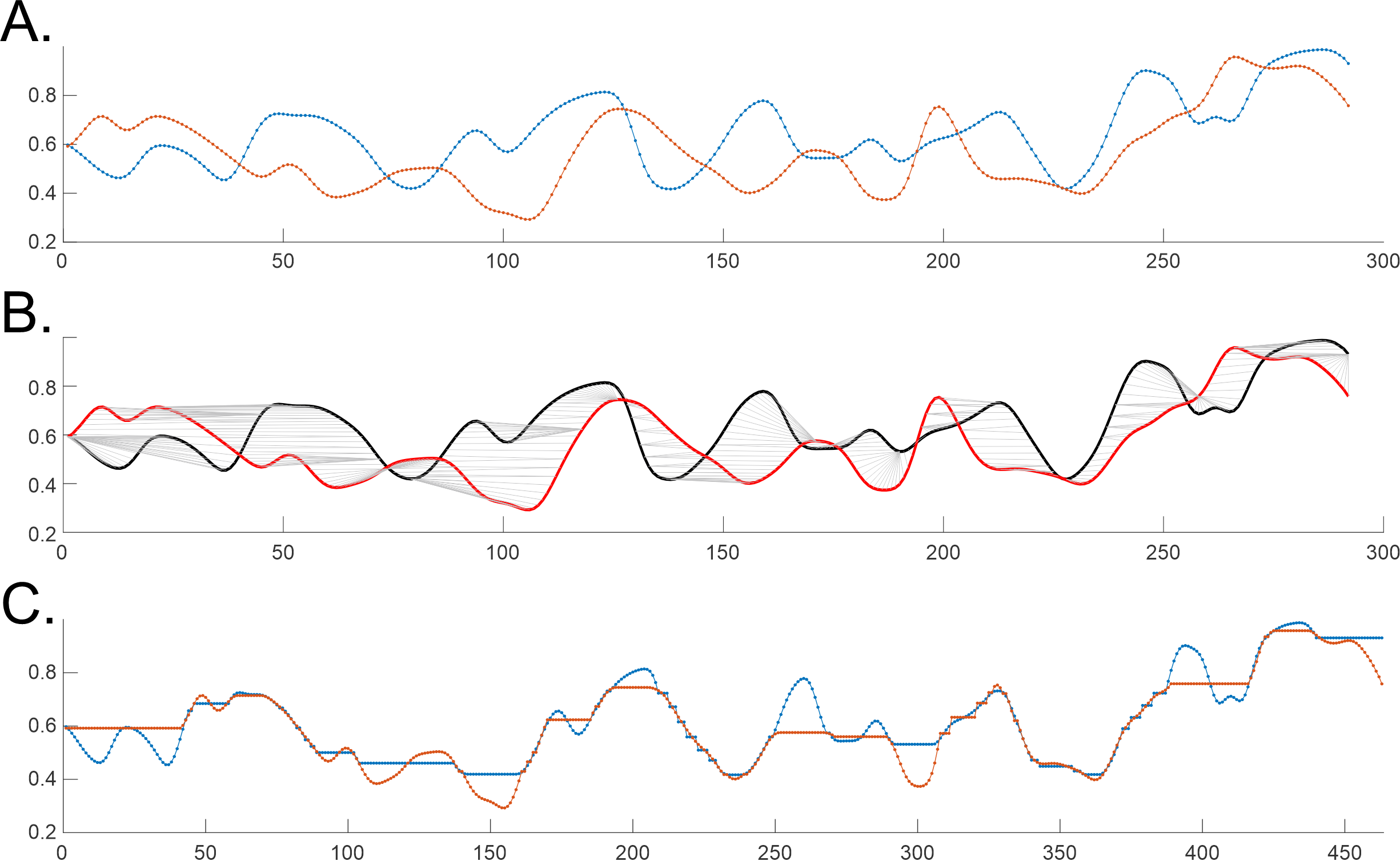}
	\caption{
		{\footnotesize 
			\textbf{Dynamic Time Warping} (DTW) method is based on applying series of stretches that maximize
			alignment between two profiles at minimal cost, without omitting elements or scrambling their order.
			DTW is commonly used to quantify similarity between waveforms, e.g., for recognizing similar speech
			patterns \cite{Sakoe}, and is naturally suitable for comparing spectral waves.
			\textbf{A}. Two concurrent segments of the hippocampal, $\nu_{\theta}^h$ ( blue line), and the
			cortical, $\nu_{\theta}^c$ (red line), spectral waves. Each segment contains about $300$ data
			points (time-wise this amounts to about $40$ ms), normalized by their respective means and shifted
			vertically into the $0\lesssim y\lesssim 1$ range. 
			\textbf{B}. Each point from $\nu_{\theta}^h$ is matched to one or more points of $\nu_{\theta
			}^c$ and vice versa, using MATLAB's \textsf{dtw} function. The gray lines show pairs of aligned 
			points. One-to-many connection lines (gray) mark the stretchings. Note that different segments of
			spectral waves alternately lag and outpace one another: the algorithm compensates these shifts to
			match the shapes.
			\textbf{C}. After the alignments, the number of points increases by $50\%$ (note the stretched-out
			$x$-axis). The point-by-point separation in the resulting alignment, measured in Euclidean metric
			and normalized to the original curve lengths serves quantifies the spectral waves' shape difference,
			which in this case amounts to $7\%$.
		}
	}
	\label{fig:dtw}
\end{figure}



\renewcommand{\figurename}{Fig.}
\begin{figure}[h]
	\includegraphics[scale=0.77]{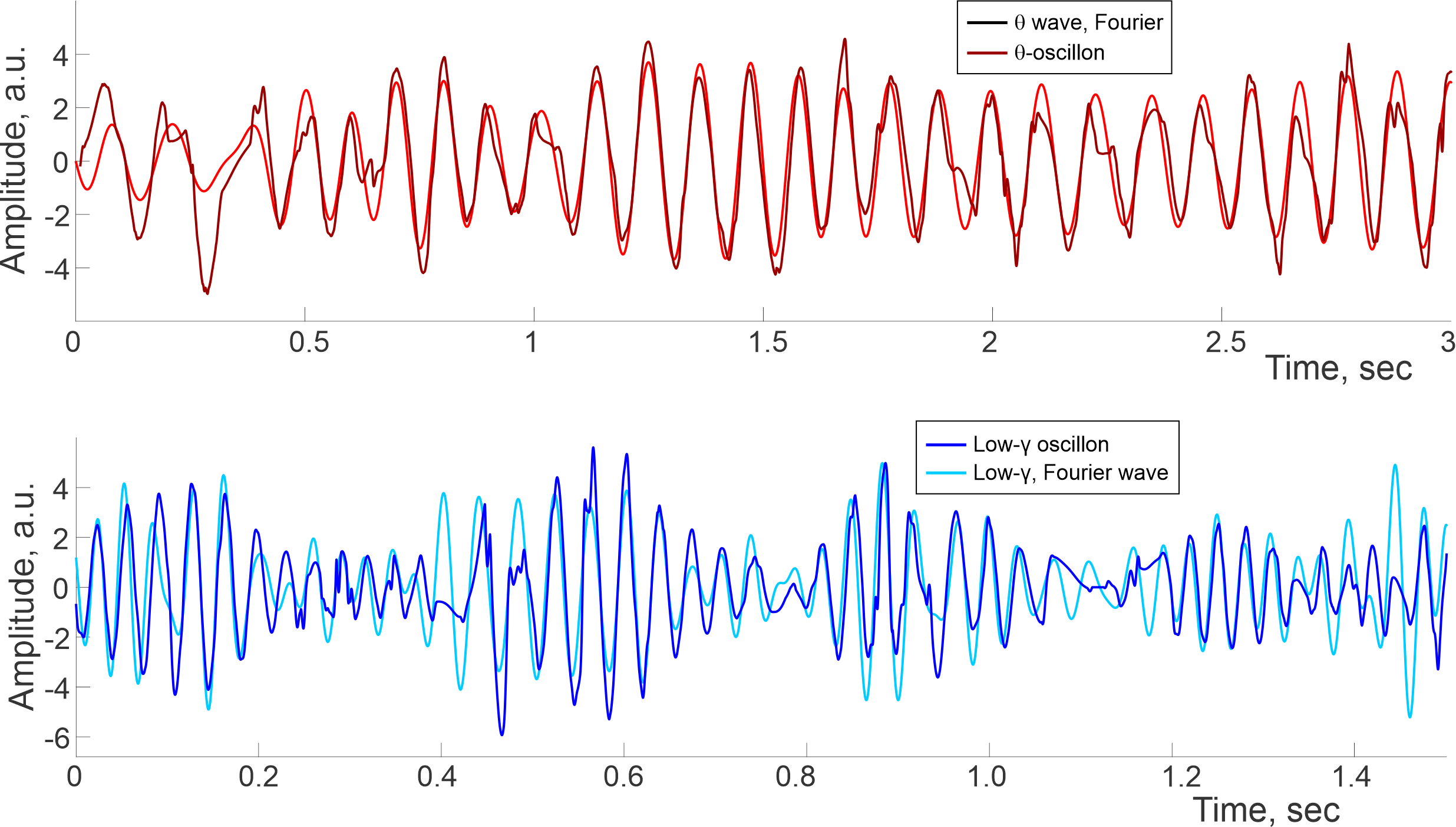}
	\caption{\footnotesize
		\textbf{Comparative waveforms} of Fourier $\theta$-wave ($4-12$ Hz) and $\theta$-oscillon (top panel)
		and	slow-$\gamma$ wave, Fourier-filtered between $20$ and $40$ Hz, compared to slow-$\gamma$ oscillon
		(bottom panel). Despite similar oscillation rates, the wave shapes are different.
	}
	\label{fig:thgm}
\end{figure} 


\renewcommand{\figurename}{Fig.}
\begin{figure}[h]
	\includegraphics[scale=0.77]{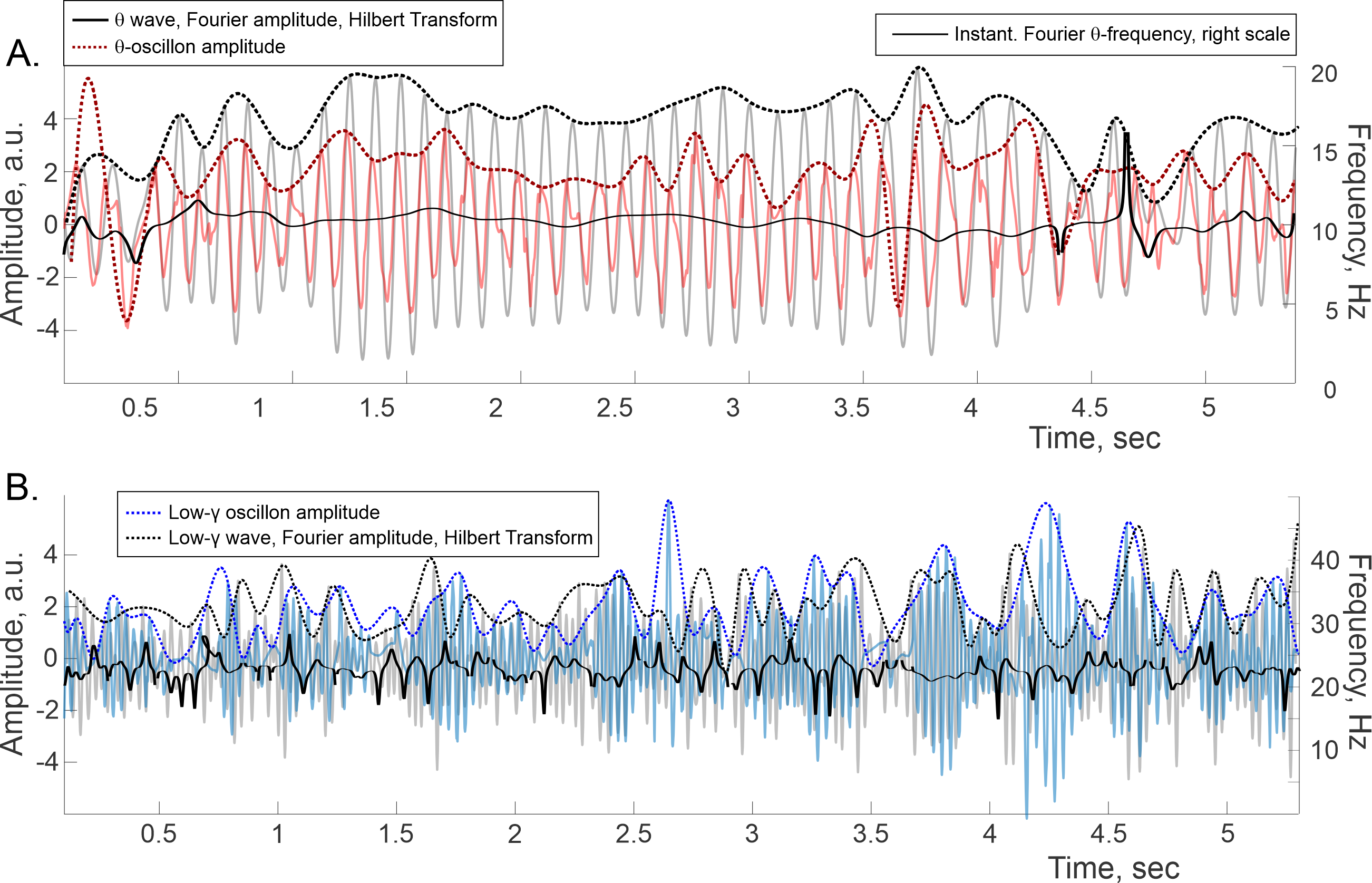}
	\caption{\footnotesize
		\textbf{Hilbert transform}. \textbf{A}. Fourier-defined $\theta$-wave (gray waveform in the background)
		shown with its amplitude (dotted line on top) and the corresponding instantaneous frequency (black line,
		placed according to the right scale) produced by Hilbert transform. Pink waveform in the foreground 
		shows $\theta$-oscillon and its amplitude computed from the net contribution of stable poles
		contributing to spectral wave (dotted brown line). 
		\textbf{B}. Fourier-defined slow-$\gamma$ wave ($20-40$ Hz), compared to the slow-$\gamma$ oscillon.
		Amplitudes and frequencies are as above. The amplitude of the latter is lower because it does not 
		include the noise component.
	}
	\label{fig:hil}
\end{figure} 



\renewcommand{\figurename}{Fig.}
\begin{figure}[h]
	\centering
	\includegraphics[scale=0.75]{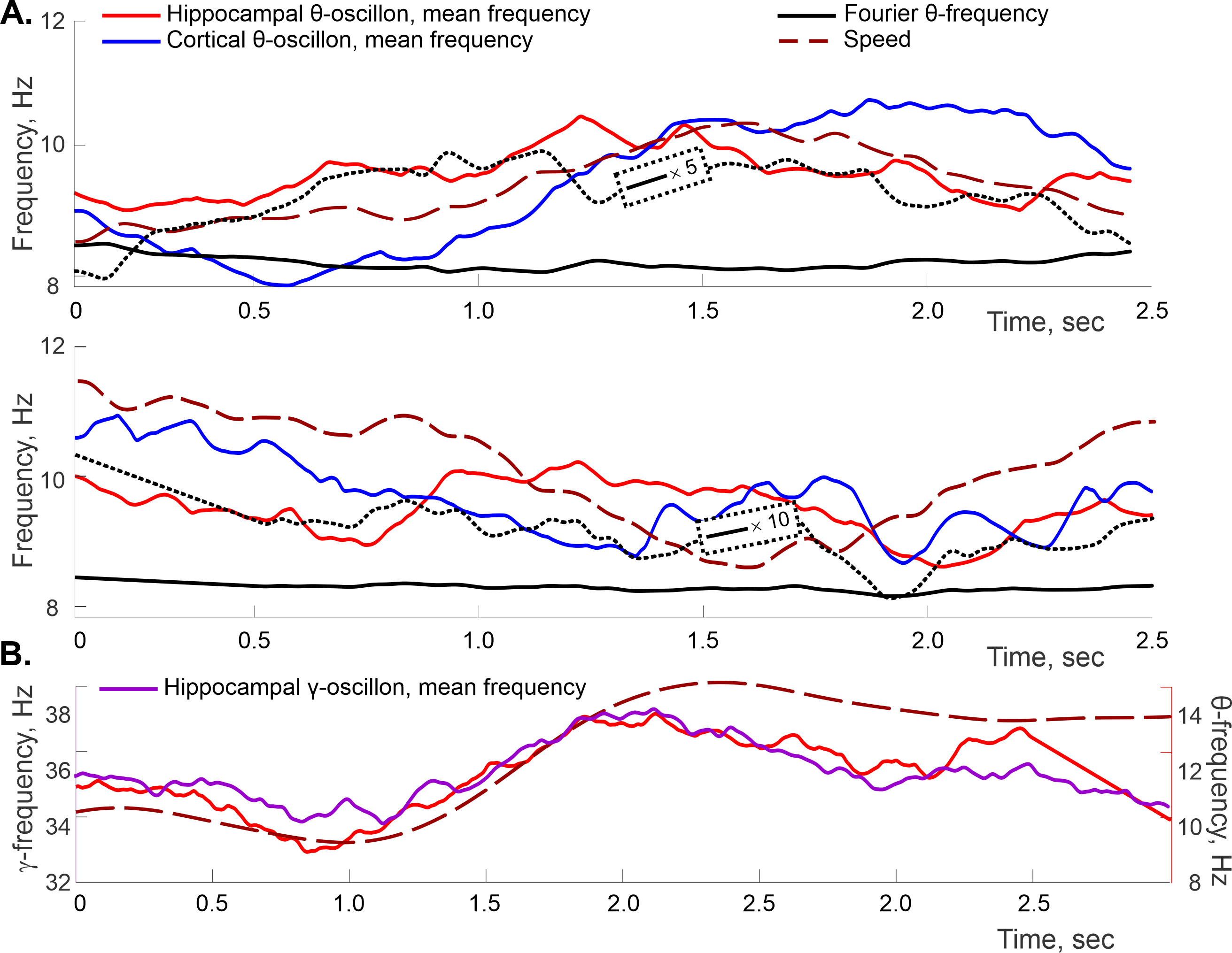}
	\caption{{\footnotesize
			\textbf{Speed vs. mean $\theta$-frequency coupling}.
			\textbf{A}. Additional examples demonstrating covariance between the moving mean of the hippocampal
			(blue) and the cortical (red) $\theta$-frequency with the rat's speed (dashed brown curve). The
			latter is scaled vertically and shifted as on Fig.~\ref{fig:spw}, to match the frequency ranges.
			The instantaneous frequency of the traditional, Fourier-defined $\theta$-waves is shown by solid
			black curve, as on Fig.~\ref{fig:mean}. A five-fold (top panel) and ten-fold (bottom panel) vertical
			stretch of the Fourier-frequency produces the dotted black curve, whose similarity to the spectral
			waves' means explains the general correspondence between our results and conventional evaluations of
			speed-frequency	couplings.
			\textbf{B}. Mean frequency profile of a hippocampal $\theta$-oscillon, left-shifted by $300$ ms
			(right scale), vs. the mean frequency of a hippocampal slow-$\gamma$ oscillon, dark
			lilac, right-shifted by $500$ ms (left scale), shown with the rat's speed profile.
	}}
	\label{fig:sp}
\end{figure}



\renewcommand{\figurename}{Fig.}
\begin{figure}[h]
	\includegraphics[scale=0.9]{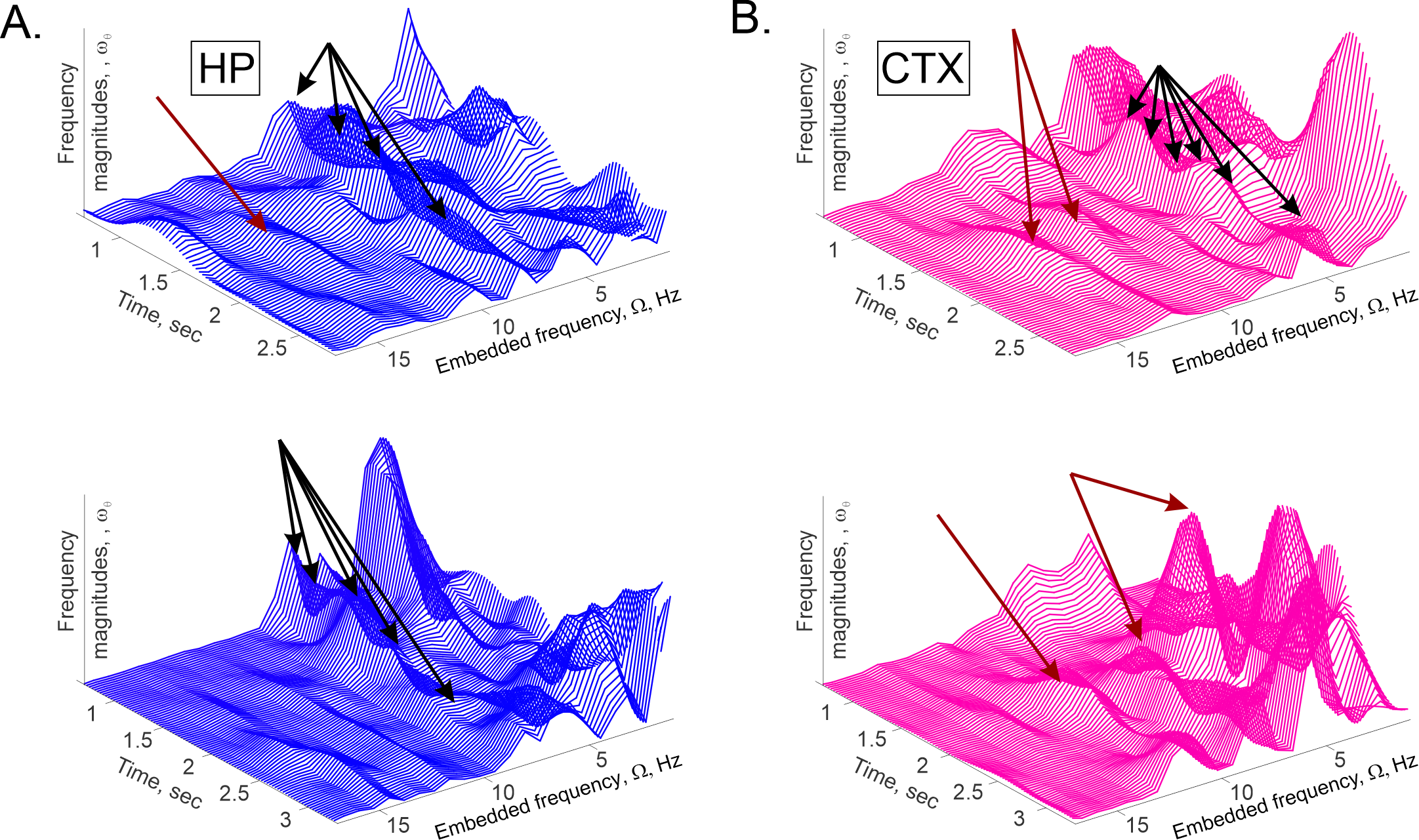}
	\caption{\footnotesize
		\textbf{Additional examples} of hippocampal (left column) and cortical (right column) $W$-spectrograms,
		illustrating the embedded frequency dynamics for the $\theta$-oscillons. Dark red arrows point at the 
		appearances of isolated	peaks and the black arrows point at the ``seedbeds" of peaks recurring at the
		same frequency.
	}
	\label{fig:w5} 
\end{figure} 



\renewcommand{\figurename}{Fig.}
\begin{figure}[h]
	\includegraphics[scale=0.77]{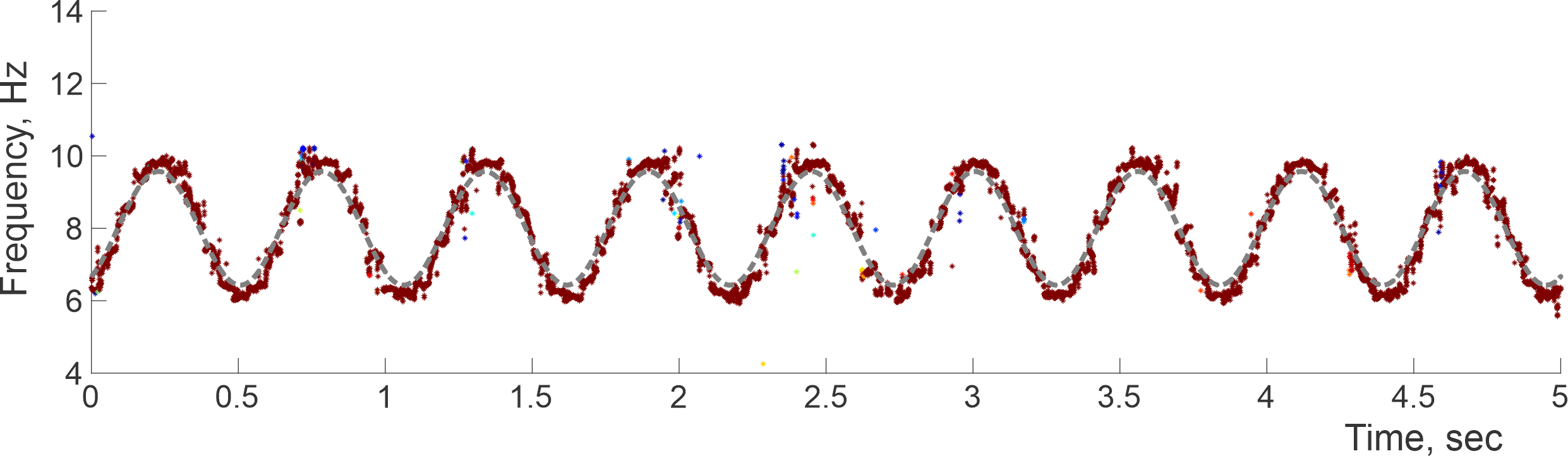}
	\caption{\footnotesize
		\textbf{Solitary spectral wave}, used to simulate the oscillatory model (\ref{kif}) with a single
		modulating frequency, $\Omega_1/2\pi\approx1.8$ Hz, oscillating around the mean $\nu_0=8$ Hz with the
		magnitude $\nu_1\approx2$ Hz (dashed gray line in the foreground). Dots represent the instantaneous
		DPT stable frequencies, colored according to their respective amplitudes. Physiological spectral waves
		shown on Figs~\ref{fig:DPT} and \ref{fig:mean} contain more than one frequency-modulating harmonic.
	}
	\label{fig:sol}
\end{figure} 



\renewcommand{\figurename}{Fig.}
\begin{figure}[h]
	\includegraphics[scale=0.9]{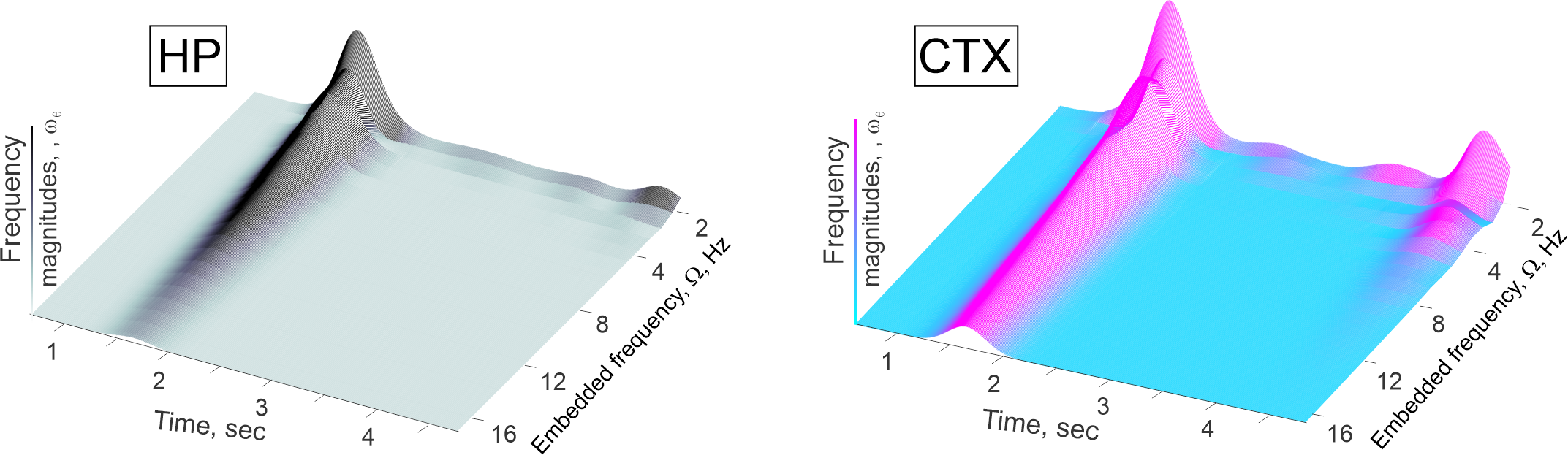}
	\caption{\footnotesize
		\textbf{Welch spectrograms of Fourier instantaneous frequencies} (solid black line on the 
		Fig.~\ref{fig:mean}B) built for the hippocampal (left) and cortical (right) spectral-$\theta$-waves,
		capture major frequency ``splashes" at about $1.7$ sec and $4.67$ sec, but do not resolve rapid
		modulations of $\theta$-frequency in-between, compare to Fig.~\ref{fig:w5}.
	}
	\label{fig:fw}
\end{figure} 


\clearpage
\newpage
\subsection{Mathematical supplement}
\label{sec:met}

\textbf{Fourier approach}\label{Four} views signals as superpositions of harmonic waves. To evaluate their
amplitudes, $2N$ data values, $\bar{\ell}_k=\{\ell_{k,1},\ell_{k,2},\ldots,\ell_{k,2N}\}$, centered around a discrete
moment $t_k$, are convolved with a set of $2N$ discrete harmonics, $e^{i\pi/N}, e^{i2\pi/N},e^{i3\pi/N},\ldots,1$,
\begin{equation}
	A_{k,N} = \sum_{n=0}^{2N-1} \ell_{k,n} z_{2N}^{-n},
	\label{ft}
\end{equation}
where $z_{2N}=e^{i\pi/N}$.
The longer is the sample set, the more precise is the spectral decomposition (\ref{ft}). Specifically, if the
original signal is a superposition of $N_P$ oscillators,
\begin{equation}
	\ell(t) = \sum_{p=1}^{P} a_p e^{i2\pi(\nu_p t + \varphi_p)},
	\label{rsig}
\end{equation}
with the amplitudes $a_p$, frequencies $\nu_p$, and phases $2\pi\varphi_p$, sampled at discrete times $t_k=k
\sigma$, then the Fourier amplitudes computed for a large number of data points are
\begin{equation}
	A_{k,l} = \sum_{k}\ell(t_k) z_{2N}^{-k} =  \sum_p\frac{a_p}{1-z_l e^{i2\pi(\nu_p\sigma+\varphi_p)}}=
	\sum_p\frac{a_p}{1-e^{i2\pi(\nu_p\sigma+\varphi_p-l/N)}}.
	\label{reg}
\end{equation}
The closer a particular frequency $\nu_p\sigma+\varphi_p+i0$ is to a $l/N$, the higher is the amplitude
$A_{k,l}$ of the corresponding harmonic. The segment selected for the analyses is then shifted, $\bar{\ell}_k
\to\bar{\ell}_{k+1}$, yielding the next value of the amplitude, $A_{k+1,l}$, and so forth, over the entire signal
span \cite{Jacobsen}.

If the signal also contains noise,
\begin{equation*}
	\ell(t) = \sum_{p=1}^{P} a_p e^{i2\pi(\nu_p t + \varphi_p)}+\xi(t),
\end{equation*}
then the Fourier peaks broaden and their heights reduce \cite{Newland}. Similar effects appear if the signal is
nonstationary, e.g., if the frequencies $\nu_p$ change with time, since it becomes more difficult to match
both the temporal and frequency details \cite{Vugt,Folland,Grunbaum}. Indeed, resolving a spectral
structure $X$ that lasts over a period $T_X$, requires using time resolution, $\Delta T$, shorter than $T_X$ 
\cite{Brigham}. On the other hand, the frequency resolution, $\Delta \nu$, should be smaller than the
$X$'s spectral size. The problem is that improving the temporal resolution (lowering $\Delta T$) reduces the
number of data points caught into the sliding window, which then lowers the frequency resolution, $\Delta\nu=1/
\Delta T$, which limits the precision of the method altogether. 

For example, the characteristic amplitude of the spectral waves shown on Fig.~\ref{fig:DPT}C and 
Fig.~\ref{fig:mean}A is about $\Delta \nu\approx 7 - 12$ Hz, which, at the Nyquist frequency $S=4$ kHz, requires
at least $N=600$ discrete harmonics, i.e., $N=600$ data points per sliding window, that can be acquired over
$\Delta T=N/S=150$ ms. On the other hand, the characteristic period of the spectral waves is about $T_{\theta}
\sim 50-100$ ms, i.e., in order to resolve the raising and the lowering phases of the spectral wave, $\Delta T$
should be less than $50$ ms. Thus, the temporal and the frequency resolutions work against each other and leave
certain details of $X$ unresolved.

\textbf{Discrete Pad\'{e} Transform}\label{Pade} (DPT) approach is based on adapting the frequencies $\nu_1,
\nu_2,\ldots,\nu_N$ according to the signal's structure, without restricting them to a regular ``frequency
grid" as in (\ref{ft}) \cite{Bessis1,Bessis2,Perotti1}. First, the discrete variable $z_{2N}$ in (\ref{ft}) and
(\ref{rsig}) is replaced with a generic, continuous complex variable $z$, thus turning the sum (\ref{rsig})
into a $z$-transform of the data series,
\begin{equation}
	S(z) = \sum_{n=1}^{\infty} s_n z^{-n}.
	\label{rz}
\end{equation}
The function (\ref{rz}) is then approximated by a ratio of two polynomials,
\begin{equation*}
	S_N(z) = P_{N-1}(z)/Q_N(z),
\end{equation*}
which constitutes the $N$-th order Pad\'e approximation, $S(z)= S_{N}(z)+o(z^{2N})$ \cite{Baker} (hence the 
name of the method \cite{Oscs1}). As shown in \cite{Bessis1,Bessis2}, the poles of $S_N(z)$, i.e., the roots of
$Q_N(z)$, capture the spectral structure of the signal $\ell(t)$, similarly to the Fourier transform (\ref{reg}). 
Specifically, for the signal (\ref{rsig}) one gets a rational function of degree $(N_P-1)/N_P$,
\begin{equation*}
	S(z) = \sum_{k} s(t_k) z^k =  \sum_p \frac{a_p e^{i\varphi_p}}{1-z e^{i2\pi\nu_p \sigma}},
\end{equation*}
with the poles $$z_p= e^{-i2\pi\nu_p\sigma},$$ and their residues defining the individual frequencies, $\nu_p$,
the amplitudes, $a_p$, and the phases, $\varphi_p$, of the corresponding oscillators.

If the signal contains a stochastic component $\xi(t)$, then the discretized time series are ``noisy," $s_n=r_n
+\xi_n$, and the generating function $S(z)$ acquires an ``irregular'' part 
\begin{equation*}
	\Xi(z) = \sum_{n=0}^{\infty} \xi_nz^{-n}.
\end{equation*}
As it turns out, the poles of the Pad\'e approximant to $\Xi(z)$ concentrate around the unit circle in the
complex plane \cite{Steinhaus} and pair up with its roots, forming the so-called Froissart doublets
\cite{Froissart,Gilewicz1,Gilewicz2,Barone}. A typical pole-zero distance in the complex plane is smaller 
than $10^{-6}-10^{-7}$ in the standard Euclidean metric, which allows detecting the Froissart doublets
numerically.
Additionally, these doublets are highly sensitive to the parameter changes, e.g., to sliding window size,
whereas the unpaired poles of (\ref{reg}) remain stable and isolated. These differences allow delineating the
LFP's noise component from the oscillations encoded by the stable poles \cite{Bessis1,Bessis2}.

\textbf{Data analyses}.
\label{Data} 
The mean amplitudes of the LFP time series was normalized to $\bar{\ell}(t) = 2$, with small amount of noise,
$\delta\xi\approx 0.01\%$ of the total amplitude, added for numerical stability. The signal was then filtered
into $1\leq f\leq60$ Hz band. The original sampling rate, $S=8$ kHz, was interpolated to $36$ kHz to improve
the low-frequency spectral wave reconstruction. We then produced $2-3$ times undersampled sub-series, which
were used for independent estimations of the regular frequencies. The sliding window width varied between
$n_{\varpi}=100$ to $n_{\varpi}=200$ (for each undersampled subseries), which yields Pad\'e approximants of
orders $N=50-200$. At the interpolated frequency,  this corresponds to $T_{\varpi}=8$ ms to $T_{\varpi}=50$
ms time windows. To ensure maximal contiguity of the spectral waves, windows were shifted by one data point.
These results remain stable under parameter variations, e.g., changes of the sliding window width \cite{Oscs1}.
The unstable frequencies were identified by detecting the Froissart doublets, with the critical pole-zero
distance $d_F=10^{-6}$ \cite{Bessis1,Bessis2,Perotti1,Oscs1}.

\textbf{Welch transform} allows estimating power spectra in transient signals \cite{Welch}. Standard power
spectra are evaluated by performing the discrete Fourier transform of the entire signal and computing the
squared magnitude of the result. In Welch's approach, the signal is first split into a large number of highly
overlapping shorter segments, and then the power spectrum of each segment is evaluated independently.

\begin{wrapfigure}{c}{0.4\textwidth}
	\includegraphics[scale=0.6]{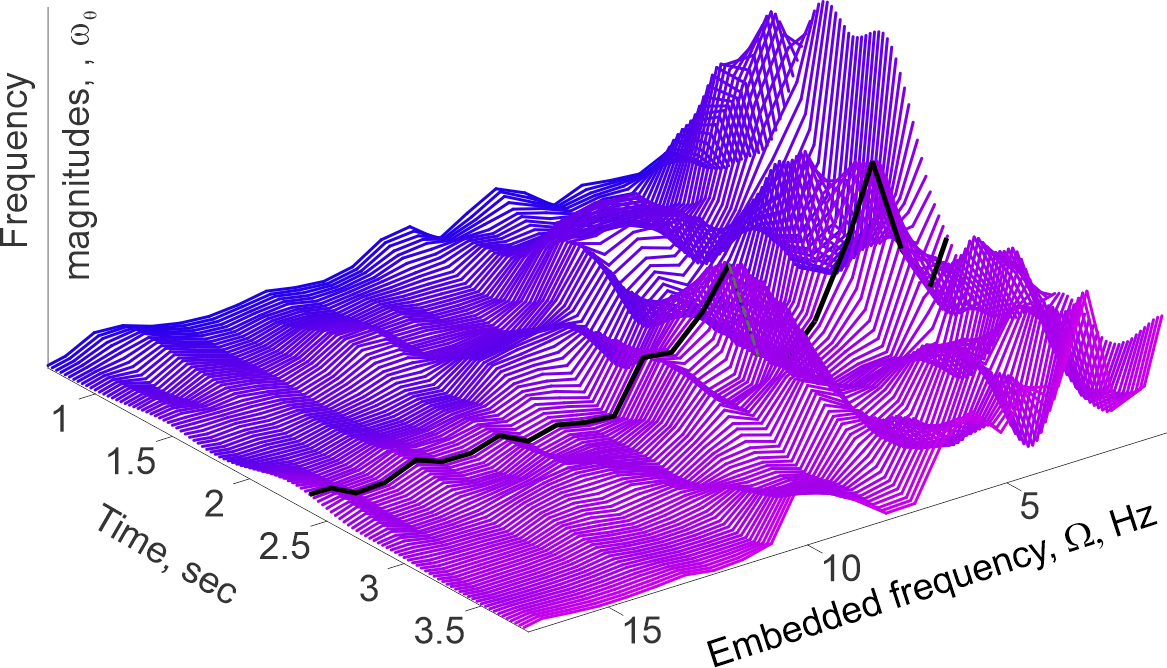}
	\caption{\footnotesize
		\textbf{Welch spectrogram} of a hippocampal spectral wave. The black line shows power profile computed
		for a particular $600$ ms long segment, centered at $2.4$ sec. The select profile shows	two peaks, which,
		over time, change their heights and positions, revealing the dynamic frequency landscape.
	}
	\label{fig:w} 
\end{wrapfigure}

The power peaks obtained from a particular data segment thus mark the most prominent frequencies appearing over
the corresponding time interval. Arranging such power profiles next to each other in natural order, one gets
three-dimensional $W$-spectrograms illustrated on Fig.~\ref{fig:w}. By construction, the lateral sections of
$W$-spectrograms are the instantaneous power---frequency profiles, whereas the longitudinal sections show the
peaks' dynamics, that highlight the evolution of the corresponding embedded frequencies, $\Omega_{\theta,i}$.

\textbf{Spectral waves}\label{SPW}. Since the poles are computed independently at each time step, based on a
finite number of data points, the patterns of reconstructed frequencies contain gaps and irregularities. To
capture the underlying continuous physical processes, we interpolated the ``raw" spectral traces over uniformly
spaced time points and used Welch transform to analyze the embedded frequencies. The mean frequency was 
evaluated as the spectral waves' moving mean, over periods comparable with largest undulation span $\sim200$
ms. All computations were performed in \textsf{MATLAB}.

\textbf{Coupling between speed and the embedded frequencies} (the dependence (4)) was obtained by evaluating
the height of peaks on $W$-spectrogram at consecutive moments of time and comparing them to the ongoing speed
values. Computations were made for peaks exceeding $20\%$ of the mean height of $W$-spectrograms, for each 
analyzed data segment. All computations were performed in \textsf{MATLAB}.

\textbf{Kuramoto model}\label{Krm} describes a network of oscillators, coupled via the equation
\begin{equation*}
	\dot{\varphi}_k=2\pi\nu_k+\frac{\lambda}{m_k}
	\sum_{l=1}^{N}C_{kl}\sin\left(\varphi_l-\varphi_k\right),
\end{equation*}
where $m_k$ is the valency of the oscillator $\ell_k$, and $C_{kl}$ is the adjacency matrix
\begin{equation*}
	C_{kl}=\left\{ 
	\begin{array}{cc}
		1\text{ if $\ell_k$ is connected to $\ell_l$}, \\ 
		0\text{ otherwise,}
	\end{array}
	\right. 
\end{equation*}
and $m_k$ is the valency of the node $\ell_k$. In this study, the network had scale-free connectivity. The mean
field produced at the $k^\mathrm{th}$ node is
\begin{equation}
	A_{K,k}e^{i\phi_K}=\sum_{l=1}^{N}C_{kl}e^{i\varphi_l}.
	\label{A}
\end{equation}
Multiplying both sides of (\ref{A}) by $e^{-i\varphi_k}$ and taking the imaginary part, one recovers the 
equation (\ref{dyn}), which highlights the mean field dynamics.

\end{document}